\documentclass[%
 reprint,
superscriptaddress,
 amsmath,amssymb,
 aps,
]{revtex4-2}
\usepackage{hyperref}
\hypersetup{colorlinks=true,citecolor=red,linkcolor=blue,urlcolor=blue}
\usepackage{amsmath}
\usepackage{xcolor}
\usepackage{enumitem}
\usepackage{graphicx}
\usepackage{dcolumn}
\usepackage{bm}

\usepackage{orcidlink}
\begin{document}

\preprint{APS/123-QED}

\title{Matter Geometry Coupling and Casimir Wormhole Geometry}

\author{A. S. Agrawal\orcidlink{0000-0003-4976-8769}}
\email{asagrawal.sbas@jspmuni.ac.in}
\affiliation{Department of Mathematics, Jayawant Shikshan Prasarak Mandal University Pune-412207, India.}
\author{Sankarsan Tarai\orcidlink{00000-0001-7209-4710}}
\email{tsankarsan87@gmail.com}
\affiliation{Department of Mathematics, School of Advanced Sciences, Vellore Institute of Technology, Chennai-600127, India}
\author{B. Mishra\orcidlink{0000-0001-5527-3565}}
\email{bivu@hyderabad.bits-pilani.ac.in}
\affiliation{Department of Mathematics, Birla Institute of Technology and Science-Pilani,\\ Hyderabad Campus, Hyderabad-500078, India.}
\author{S.K. Tripathy\orcidlink{0000-0001-5154-2297}}
\email{tripathy\_sunil@rediffmail.com}
\affiliation{Department of Physics, Indira Gandhi Institute of Technology, Sarang, Dhenkanal, Odisha-759146, India.}

\date{\today}

\begin{abstract}
In this study, we investigate traversable wormhole solutions within the set up of $f(R,\mathcal{L}_{m})$ gravity, a modified theory of gravity where the gravitational action relies upon the the matter Lagrangian $\mathcal{L}_{m}$ and the Ricci scalar $R$. In General Relativity (GR), stability issue in traversable wormholes  necessitates the existence of exotic matter violating the null energy condition (NEC). In contrast, we explore wormhole solutions that align with the criteria for Casimir wormholes, which do not necessarily require NEC violation. Our analysis demonstrates that in the context of $f(R,\mathcal{L}_{m})$ gravity, exotic matter can sustain these wormholes. We further examine the traversability conditions of the wormhole, considering both scenarios with and without the Generalized Uncertainty Principle (GUP) correction. Additionally, the stability of the wormhole is assessed based on equilibrium conditions. Our findings suggest that $f(R,\mathcal{L}_{m})$ gravity offers a viable framework for the existence of stable, traversable wormholes sustained by exotic matter, potentially expanding the landscape of viable wormhole solutions beyond the confines of GR.

\end{abstract}

\maketitle

\section{Introduction}
Wormholes conceived as a hypothetical tunnel joining two points in space-time has a subject of interest in the sense of fun for a dream travel through a shortcut distance and time. Besides being a subject matter of fiction for a long time, realising traversable wormholes with usual fluid has posed a considerable challenge to theoretical physicists. The pioneering investigation into traversable wormholes with substantial stability was conducted by Morris and Thorne \cite{Morris95}. Their investigation delved into the potential for human time travel within the framework of General Relativity (GR). According to GR, the presence of matter results in the flexibility and deformability of space-time, as opposed to rigidity. Furthermore, the more massive an entity, the more pronounced the curvature of space, thereby leading to the singularity and formation of black holes. In the later scenario, there is a break down of the texture of the space-time. However, should the formation of the singularity be circumvented, the possibility of traversing the throat arises, thereby removing constraints on the observer's movement within the manifold. After the discovery of GR, Flamm \cite{flamm1916} was the first to examine the probability of a solution to the Einstein field equations. However, it was later shown that his solution was unstable. Following the work of Flamm, Einstein, and Rosen, a more precise understanding of the wormhole structure was developed, leading to the well-known concept of the Einstein–Rosen bridge \cite{Einstein73}.

Over and above, many researchers have put forth several extended theories of gravity to account for the much-discussed late-time cosmic speed-up issue and the associated dark energy concept \cite{Peebles2003}. Geometrically modified theories of gravity, serving as suitable generalized extensions of GR, involve the modification of the geometric action to account for dark energy and provide a comprehensive explanation for the early and late-time accelerated expansion of the Universe. Modified gravity theories significantly contribute to geometrical action and have effectively addressed various astrophysical issues concerning wormholes and compact stars. A good number of research works on wormhole solutions are documented in the literature within the context of diverse modified gravity theories, including $f(R)$\cite{Lobo09, Rahaman14, Maz16, GODANI20, Mishra21, Ghosh21, Agrawal22}, Gauss-Bonnet\cite{Bhawal92, Jusufi20, ZHANG23}, $f(R, T )$ theory \cite{ban2021, Ilyas2022, GODANI2022, Gashti23}, teleparallel \cite{Jamil13, Sharif13, Tefo18}, symmetric teleparallel \cite{Farook10, Ban21, Kiro23}, modified teleparallel gravity\cite{Bhmer2012}, brane \cite{Luis62} and  Rastall gravity \cite{Naza2023, Halder2019, Heyd2023}. Furthermore, for evaluating the nature of the Universe, $f(R)$ gravity has established a dependable framework. The $f(R, \mathcal{L}_{m})$ gravity theory \cite{Harko:2010:70} extends the $f(R)$ framework by expressing the  Lagrangian as an arbitrary function of both $R$, and $\mathcal{L}_{m}$. This theory seeks to serve as a generalization of both geometry and matter within the context of gravitational interactions.

A wormhole is theorized as a topological feature of spacetime, resembling a tube-like structure that is asymptotically flat at both ends. The wormhole throat radius may either remain constant or vary. If a wormhole remains unchanged over time, it is called a static wormhole. If it changes, it is called a non-static wormhole. A fundamental requirement for the formation of a stable wormhole within the framework of GR is the presence of exotic matter, which violates the null energy condition (NEC) \cite{Visser1995, Hochberg21, Morris95}. In contrast, usual matter content fulfills this energy condition \cite{Samanta39, Harko04}.  
In quantum physics, it is well-established that violations of the NEC can be readily achieved in certain directions \cite{Wyman39}. Consequently, much of the research on wormhole physics centers on semi-classical gravity, where the quantum energy-momentum tensor is regarded as a potential source of the gravitational field \cite{Sushkov50}. Notably, in the case of a scalar field conformally coupled to gravity, the stress-energy tensor can violate the NEC even within classical frameworks \cite{Wald33}. At higher energy scales, though still well below the Planck scale, there may be additional forms of classical violations of the NEC. These include higher derivative theories \cite{Callan42, Hochberg49}, Brans–Dicke theory \cite{Soma07, Camera11, Trobo15}, and potentially even more exotic theoretical frameworks.

In 1948, Hendrik Casimir first introduced the concept of the Casimir effect \cite{casimir1948}, suggesting that two uncharged, parallel plates placed nearby within a vacuum would experience an attractive force arising from quantum fluctuations rather than any classical interaction. More recently, an experiment conducted by Lamoreaux \cite{Lamo1997} confirmed the existence of this phenomenon. According to quantum electrodynamics, the quantum fluctuations of the vacuum between the plates can generate a negative energy density, which has been proposed as a potential energy source for the construction of traversable wormholes. Garattini \cite{Garattini2019} suggested that Casimir energy could be a pathway for investigating traversable wormholes and examined how weak energy conditions affect the possibility of traversing through the wormhole. 

With a motivation to construct stable traversable wormholes with a cosmic fluid satisfying the NEC, in the present work, we worked within the ambit of a geometrically modified gravity theory dubbed as the $f(R,L_m)$ gravity. We explored the negative energy density source similar to the Casimir effect within the wormhole geometry and incorporated the GUP correction to observe its effect on the energy conditions. The paper is structured as follows: In Section \ref{sec:cwfrl}, we introduce the  $f(R,\mathcal{L}_{m})$ modified gravity in the context of traversable wormhole conditions, laying the foundation for our study. Section \ref{CE} is dedicated to exploring the wormhole solution considering the energy source from the Casimir effect. In Section \ref{WG}, we delve into the geometric structure of the wormhole, providing detailed insights into its spatial properties. The stability of the wormhole solution is examined in Section \ref{SC}, where we discuss its equilibrium conditions to assess its viability. Finally, in Section \ref{Conclusion}, we summarize our findings.

\section{Wormholes in the \texorpdfstring{$f(R,\mathcal{L}_m)$}{} gravity}\label{sec:cwfrl}

We consider the action for the $f(R,\mathcal{L}_m)$ gravity as \cite{Harko:2010:70}

\begin{equation}\label{cfrl1}
    S=\int d^4x \sqrt{-g}f(R,\mathcal{L}_m).
\end{equation}
The Ricci tensor, denoted as $R_{\mu \nu}$, plays a crucial role in representing the curvature of spacetime, and the Ricci scalar is the contracting form of the Ricci tensor defined as $R=R_{\mu \nu}g^{\mu \nu}$. The metric tensor, represented by $g_{\mu \nu}$, defines the geometric structure of the space. The expression $g$ denotes the metric determinant and is assumed to be applicable in a unit system where $8\pi G=c=1$.

By utilizing the variational principle in the aforementioned action, one can derive the field equations for $f(R,\mathcal{L}_m)$ gravity as follows:

\begin{eqnarray}\label{Eq:field1}
    R_{\mu\nu}f_R(R,\mathcal{L}_m)+(g_{\mu\nu}\nabla_\mu\nabla^\mu-\nabla_\mu\nabla_\nu )f_R(R,\mathcal{L}_m)\nonumber \\ 
    -\frac{1}{2}f(R,\mathcal{L}_m)g_{\mu\nu}=\frac{1}{2}f_{\mathcal{L}_m}(R,\mathcal{L}_m)(T_{\mu\nu}-\mathcal{L}_mg_{\mu\nu})\,.\label{cfrl2}
\end{eqnarray}

The function $f(R,\mathcal{L}_{m})$ is pivotal in this context and is expressed as a combination of two parts: $f_{1}(R)$ and 
$f_{2}(R)G(\mathcal{L}_{m})$. Here, $f_{1}(R)$ and $f_{2}(R)$ are arbitrary functions depending on the Ricci scalar $R$, which encapsulates the curvature of spacetime. $G(\mathcal{L}_{m})$, on the other hand, is a function of the matter Lagrangian density $\mathcal{L}_{m}$, which accounts for the distribution of matter within the spacetime. The partial derivatives 
$f_{R}(R,\mathcal{L}_m)$ and $f_{\mathcal{L}_m}(R,\mathcal{L}_m)$ represent the rate of change of $f(R,\mathcal{L}_m)$ with respect to $R$ and $\mathcal{L}_m$, respectively. These derivatives are essential in determining how the function $f(R,\mathcal{L}_m)$ responds to variations in the curvature and the matter content of the universe.

The contracting form of the field equation \eqref{Eq:field1} can be written as
\begin{eqnarray}\label{Eq:field2}
f_{R}R+3\Box f_{R}-2f=f_{\mathcal{L}_{m}}\left(\frac{1}{2}T-2\mathcal{L}_{m}\right).    
\end{eqnarray}
By eliminating the term $\Box f_{R}$ between equation \eqref{Eq:field1} and \eqref{Eq:field2} we obtain
\begin{eqnarray}
f_{R}\left[R_{\mu \nu}-\frac{R}{3} g_{\mu \nu}\right]+\frac{f}{6}-\nabla_{\mu}\nabla_{\nu}f_{R}\quad \quad \quad\nonumber \\
=\frac{f_{\mathcal{L}_{m}}}{2}\left[T_{\mu \nu}-\frac{1}{3}(T-\mathcal{L}_{m}) g_{\mu \nu} \right]\,. \label{Eq:field3}  \end{eqnarray}

The covariant divergence of equation \eqref{Eq:field1}, with the identity
\begin{eqnarray}
\nabla^{\mu}\bigg[R_{\mu \nu }f_{R}+( g_{\mu\nu}\Box-\nabla_{\mu}\nabla_{\nu})f_{R}-\frac{1}{2}fg_{\mu \nu}\bigg]=0 \,,   
\end{eqnarray}
leads to the divergence of the energy-momentum tensor $T_{\mu \nu}$ as
\begin{eqnarray}
\nabla^{\mu}T_{\mu \nu}&=& 2\nabla^{\mu}\ln[f_{\mathcal{L}_{m}}]\frac{\partial \mathcal{L}_{m}}{\partial g^{\mu \nu}}.
\end{eqnarray}

The energy-momentum conservation of the matter field, $\nabla^{\mu}T_{\mu \nu}=0$, establishes a functional relationship between the Lagrangian density and matter and consequently we get
\begin{equation}\label{cfrl3}
	\nabla^\mu\ln f_{L_m}=0\,.
\end{equation}
To derive the cosmological wormhole solution, we begin by recalling that the static wormhole metric is described by the Morris-Thorne solution \cite{Morris95}. This solution provides the foundational framework for constructing traversable wormholes, characterized by specific conditions on the redshift and shape functions to ensure stability and traversability.

\begin{equation}\label{MTM}
    ds^2=-e^{2\Phi(r)}dt^2+\frac{dr^2}{1-\frac{b(r)}{r}}+r^2(d\theta^2+\sin^2\theta d\phi^2),
\end{equation}
where $\Phi(r)$ represents the redshift function and $b(r)$ denote the shape function. For the spatial geometry of the wormhole to attain the appropriate asymptotically flat limit, the redshift function $\Phi$ must satisfy the condition that it approaches a constant value as the radial coordinate tends to infinity. This ensures the absence of gravitational time dilation at large distances from the wormhole throat.
\begin{equation}\label{RF_I}
    \lim_{r\rightarrow{\infty}}\Phi<\infty.
\end{equation}
At the wormhole throat, denoted by $r_0$, the traversable wormhole must satisfy specific conditions to ensure stability and traversability. These include the requirement that the shape function $b(r)$ equals $r_0$ at the throat, ensuring a smooth transition through the throat and that the flare-out condition is met indicating that the geometry expands outward from the throat.
\begin{subequations}
\begin{eqnarray}
    b(r_0)=r_0\,,\label{SF_I} \\
    b'(r_0)\leq 1\,,\label{SF_II}\\
    b'<\frac{b(r)}{r}\,\label{SF_III}\,.
\end{eqnarray}    
\end{subequations}
In this paper, a prime denotes differentiation with respect to the radial coordinate. To satisfy the necessary geometric conditions for a traversable wormhole, it is required that $b'(r) < {b(r)}/{r}$ at all points. This ensures that the shape function maintains the appropriate curvature for wormhole stability and traversability.

To ensure the traversability of the wormhole, it is crucial that the condition $b(r) < r$ is satisfied at all points away from the throat, $r_0$. The exotic matter sustaining the wormhole is characterized by the presence of anisotropic fluid, which violates the standard energy conditions necessary to maintain the geometry of the wormhole.

The anisotropic fluid of the wormhole is considered through the energy-momentum tensor
\begin{equation}\label{EMT_I}
T_{\mu \nu}=(\rho +p_{t})u_{\mu}u_{\nu}+p_{t}g_{\mu \nu}+(p_{r}-p_{t})x_{\mu}x_{\nu}\,,    
\end{equation}
where $u_{\mu}$ and $x_{\mu}$ respectively represent the four velocity vectors along the transverse and radial directions. They satisfy the relations $u_{\mu}u_{\mu}=-1$ and $x_{\mu}x_{\mu}=1$. Here, $\rho$ represents the energy density, while $p_{r}$ and $p_{t}$ denote the radial and tangential components of the pressure, respectively. The trace of the energy-momentum tensor is expressed as $T = -\rho + p_{r} + 2p_{t}$, providing a key relationship between the energy density and pressure components within the system.

Finally, the energy-momentum tensor for a cosmological wormhole can be expressed as:
\begin{equation}\label{EMT_II}
    T^\mu_\nu=\texttt{diag}[-\rho,p_r,p_t,p_t],  
\end{equation}
In this context, $\rho$ represents the matter-energy density of the wormhole, $p_r$ denotes its radial pressure, and $p_t$ corresponds to its tangential pressure \cite{Bahamonde2016}.

The energy density, radial and tangential pressure for the given wormhole geometry may be obtained for a generic functional form in the $f(R,\mathcal{L}_m)$ gravity theory as
\begin{subequations}
\begin{eqnarray}
\rho&=& \frac{f}{f_{\mathcal{L}_{m}}}-\frac{f_{R}}{r^{2}f_{\mathcal{L}_{m}}}\bigg[\big(r(b'-4)+3b\big)\Phi'-2r(r-b)\Phi'^{2} \nonumber \\
&&+2r(b-r)\Phi''\bigg]-\frac{f_{RR}}{r^{2}f_{\mathcal{L}_{m}}}\bigg[2r(r-b)R''-\nonumber \\
&&\big(r(b'-4)+3b\big)R'\bigg] -\frac{2f_{RRR}R'^{2}}{f_{\mathcal{L}_{m}}}\left(1-\frac{b}{r}\right)-\mathcal{L}_{m}\,,\nonumber \\
\\
p_{r} &=& - \frac{f}{f_{\mathcal{L}_{m}}} +\frac{bf_{R}}{r^{3}f_{\mathcal{L}_{m}}}\left(2r^{2}\Phi'^{2}+2r^{2}\Phi''-r\Phi'-2\right)\nonumber \\
&&+\frac{f_{R}}{r^{2}f_{\mathcal{L}_{m}}}\bigg(b'(r\Phi'+2)-2r^{2}(\Phi'^{2}+\Phi'')\bigg)\nonumber \\
&&+\frac{2f_{RR}R'}{f_{\mathcal{L}_{m}}}\left(1-\frac{b}{r}\right)\left(\Phi'+\frac{2}{r}\right)+\mathcal{L}_{m}\,, \\
p_{t}&=& -\frac{f}{f_{\mathcal{L}_{m}}}+\frac{f_{R}}{r^{3}f_{\mathcal{L}_{m}}}\bigg(b(2r\Phi'+1)+(b'-2r\Phi')r\bigg)\nonumber \\
&&+\frac{2f_{RR}}{r^{2}f_{\mathcal{L}_{m}}}\bigg[\frac{R'}{2}\big(r(2r\Phi'-b'+2)-b(2r\Phi'+1)\big)\nonumber \\
&&+r(r-b)R''\bigg] +\frac{2R'^{2}f_{RRR}}{f_{\mathcal{L}_{m}}}\left(1-\frac{b}{r}\right)+\mathcal{L}_{m}\,.
\end{eqnarray}
\end{subequations}
For a tideless wormhole, where $\Phi$ is constant, the field equations for the wormhole metric, with $R={2b'(r)}/{r^{2}}$ in the context of extended gravity, can be derived as follows:
\begin{subequations}
\begin{eqnarray}
\rho&=& \frac{f}{f_{\mathcal{L}_{m}}}-\frac{f_{RR}}{r^{2}f_{\mathcal{L}_{m}}}\bigg[(r-b)\frac{24b'+4r(b^{(3)}r-4b'')}{r^3}\nonumber \\
&&-\big(r(b'-4)+3b\big)\frac{2 \left(r b''-2 b'\right)}{r^3}\bigg]\nonumber \\
&&-\frac{8f_{RRR}}{f_{\mathcal{L}_{m}}}\frac{(r b''-2 b')^{2}}{r^6}\left(1-\frac{b}{r}\right)-\mathcal{L}_{m}\,, \label{rho_a} \\
p_{r} &=& - \frac{f}{f_{\mathcal{L}_{m}}} -\frac{2bf_{R}}{r^{3}f_{\mathcal{L}_{m}}}+\frac{2b'f_{R}}{r^{2}f_{\mathcal{L}_{m}}}\nonumber \\
&&+\frac{8f_{RR}}{f_{\mathcal{L}_{m}}}\frac{ \left(r b''-2 b'\right)}{r^4}\left(1-\frac{b}{r}\right) +\mathcal{L}_{m}\,,\label{pr_a} \\
p_{t}&=& -\frac{f}{f_{\mathcal{L}_{m}}}+\frac{f_{R}}{r^{3}f_{\mathcal{L}_{m}}}(b+rb')+\frac{2f_{RR}}{r^{2}f_{\mathcal{L}_{m}}}\bigg[\frac{(r b''-2 b')}{r^3}\nonumber \\
&&\times\big(r(2-b')-b\big)+(r-b)\frac{12b'+2r(b^{(3)}r-4b'')}{r^3}\bigg]\nonumber \\
&& +\frac{8f_{RRR}}{f_{\mathcal{L}_{m}}}\frac{(r b''-2 b')^{2}}{r^6}\left(1-\frac{b}{r}\right)+\mathcal{L}_{m}\,, \label{pt_a}
\end{eqnarray}   
\end{subequations}

$b'$, $b''$ and $b^{(3)}$ respectively denotes $\frac{db}{dr}$, $\frac{d^{2}b}{dr^{2}}$ and $\frac{d^{3}b}{dr^{3}}$.

\subsection{The Basic Formalism for the \texorpdfstring{$f(R,\mathcal{L}_{m})$}{} Model}
We consider
\begin{equation}
    f(R,\mathcal{L}_{m})=f_{1}(R)+[1+\lambda f_{2}(R)]\mathcal{L}_{m},
\end{equation}
with $f_{1}(R)=\frac{R}{2}$ and $f_{2}(R)=R$. $\lambda$ is a model parameter to be fixed from some plausible physical basis.

From Eqs. \eqref{rho_a}, \eqref{pr_a} and \eqref{pt_a}, the energy density, radial pressure, and tangential pressure can be written as
\begin{subequations}
\begin{eqnarray}
\rho&=& \frac{R}{2 \lambda  R+2}\,, \label{rho1}\\
p_{r}&=& -\frac{-2 b' (2 \lambda  \mathcal{L}_{m} r+r)+b (4 \lambda  \mathcal{L}_{m}+2)+r^3 R}{2 r^3 (\lambda  R+1)}\,, \\
p_{t}&=&  \frac{r (2 \lambda  \mathcal{L}_{m}+1) b'+2 \lambda  \mathcal{L}_{m} b+b-r^3 R}{2 r^3 (\lambda  R+1)}\,.
\end{eqnarray}    
\end{subequations}
The matter Lagrangian, $\mathcal{L}_{m}$, is a general function that depends on the matter-energy density $\rho$ and pressure $p$. Additionally, it may be associated with other thermodynamic variables such as specific entropy $s$ or baryon number $n$. Thus, the Lagrangian can be expressed as $\mathcal{L}_{m} = \mathcal{L}_{m}(\rho, p, s, n, \dots)$.

In the upcoming discussion, we will examine the cosmological consequences of using $\mathcal{L}_{m} = -\rho$ as the Lagrangian so that we have 

\begin{subequations}
\begin{eqnarray}
\rho&=& \frac{R}{2 \lambda  R+2}\,, \label{rho1a}\\
p_{r}&=& -\frac{-2 r b'(r)+2 b(r)+r^3 R (\lambda  R+1)}{2 r^3 (\lambda  R+1)^2}\,, \\
p_{t}&=&\frac{r b'(r)+b(r)-r^3 R (\lambda  R+1)}{2 r^3 (\lambda  R+1)^2}\,.
\end{eqnarray}    
\end{subequations}
\section{Wormhole geometry exploring the Casimir effect}\label{CE}

The Casimir effect is a quantum phenomenon characterized by an attractive force between two closely positioned uncharged conducting plates in a vacuum, arising due to fluctuations in the vacuum state of the electromagnetic field. Initially proposed by H. Casimir in 1948 \cite{casimir1948}, this effect was later confirmed through experimental observations \cite{sparnaay1957}. Notably, the Casimir effect is contingent on the boundary conditions of the geometry of the system. It represents a unique laboratory-realizable source of exotic matter, distinguishing it as a significant manifestation of quantum field theory.
The Casimir effect holds significant potential for applications in the study of traversable wormholes involving exotic matter. In this context, Garattini proposed a model of traversable wormholes based on the equation of state derived from the Casimir effect, commonly referred to as Casimir wormholes \cite{Garattini2019}. Recent advancements in quantum mechanics have introduced the notion of a minimal length scale, identified as the Planck length, which imposes fundamental limits on the precision of measurements at extremely small distances in spacetime.
This concept originates from quantum gravity theories and calls for a revision of the conventional position-momentum uncertainty relation, leading to the formulation of the GUP. The GUP further redefines the negative Casimir energy density, contingent upon the framework of maximally localized quantum states. Researchers are actively investigating this by applying the GUP to modify the Casimir energy density, focusing on its implications for GUP-corrected wormholes and their traversability. The exploration of the impact of GUP on modeling traversable wormholes presents a compelling area of study \cite{Tripathy2021}.

\subsection{Casimir wormholes without GUP correction}\label{withoutGUP}
The Casimir effect is characterized by an attractive force between two plates, which is attributed to the presence of renormalized negative energy. This phenomenon arises from the vacuum fluctuations of the quantum field between the plates, leading to a measurable force as a result of the constrained modes of the field.
\begin{equation}
    E(a)=-\frac{\pi^{2}}{720}\frac{S}{a^{3}}
\end{equation}
The Casimir energy decreases as the separation distance between the plates, denoted by $a$, is reduced and as the surface area of the plates, $S$, is adjusted. Consequently, this leads to a corresponding reduction in the energy density, expressed as

\begin{equation}
    \rho(a)=-\frac{\pi^{2}}{720}\frac{1}{a^{4}}\label{casimir_rho}
\end{equation}
The pressure can be determined by utilizing the renormalized negative energy, which arises from the vacuum fluctuations in the quantum field. This approach provides a precise measure of the force exerted between the boundaries in systems governed by quantum effects.
\begin{equation}
    p(a)=-\frac{1}{\rho}\frac{dE(a)}{da}=-\frac{\pi^{2}}{240}\frac{1}{a^{4}}\label{casimir_pressure}
\end{equation}
The energy density \eqref{casimir_rho} and pressure \eqref{casimir_pressure} expressions derived above give rise to an equation of state (EoS) of the form $p = \alpha \rho$, where $\alpha$ represents the EoS parameter. The Casimir force, defined as the product of the surface area and the pressure, is expressed as $F = -\frac{\pi^{2}S}{240}\frac{1}{a^{4}}$. The negative sign of this force indicates its attractive nature.

We wish to explore the Casimir effect to construct some traversable wormholes within the discussed matter-geometry coupling gravity theory. In this process, we may consider the plate distance $a$ either  as a constant parameter or to be replaced by a variable radial coordinate. Both the cases have their own importance and may lead to viable Casimir  wormhole solutions.
\subsubsection{\texorpdfstring{$a$}{} is replaced by a radial coordinate $r$}\label{caseI}

For the present case, a simplified form of the shape function can be derived by integrating Eq.\eqref{rho1a} along with the condition \eqref{casimir_rho}. The only replacement made here is considering a radial coordinate $r$ for the plate distance $a$. Since $R=\frac{2b'}{r^{2}}$, we obtain the shape function of the Casimir wormhole as
\begin{eqnarray}
    b(r)&=&\frac{\pi}{8 \sqrt{\lambda } \mu }  \bigg[\log \left(\nu  r^2+\pi  \sqrt{\lambda } (\mu  r+1)\right)\nonumber \\
    &&-\log \left(\nu  r^2+\pi  \sqrt{\lambda } (1-\mu  r)\right)-2 \tan ^{-1}(\mu  r+1)\nonumber \\
    &&+2 \tan ^{-1}(1-\mu  r)\bigg]+c_1 \label{b(r)_1}
\end{eqnarray}
where $\mu = \frac{2 \sqrt{3} \sqrt[4]{10}}{\sqrt{\pi } \sqrt[4]{\lambda }}$ and $\nu =6 \sqrt{10}$. Also,
\begin{eqnarray}
    c_{1}&=&r_{0}-\frac{\pi}{8 \sqrt{\lambda } \mu }  \bigg[\log \left(\nu  r_{0}^2+\pi  \sqrt{\lambda } (\mu  r_{0}+1)\right)\nonumber \\
    &&-\log \left(\nu  r_{0}^2+\pi  \sqrt{\lambda } (1-\mu  r_{0})\right)-2 \tan ^{-1}(\mu  r_{0}+1)\nonumber \\
    &&+2 \tan ^{-1}(1-\mu  r_{0})\bigg]\,. \label{b(1)_2}
\end{eqnarray}
\begin{figure}[ht]
    \centering
    \includegraphics[width=0.45\textwidth]{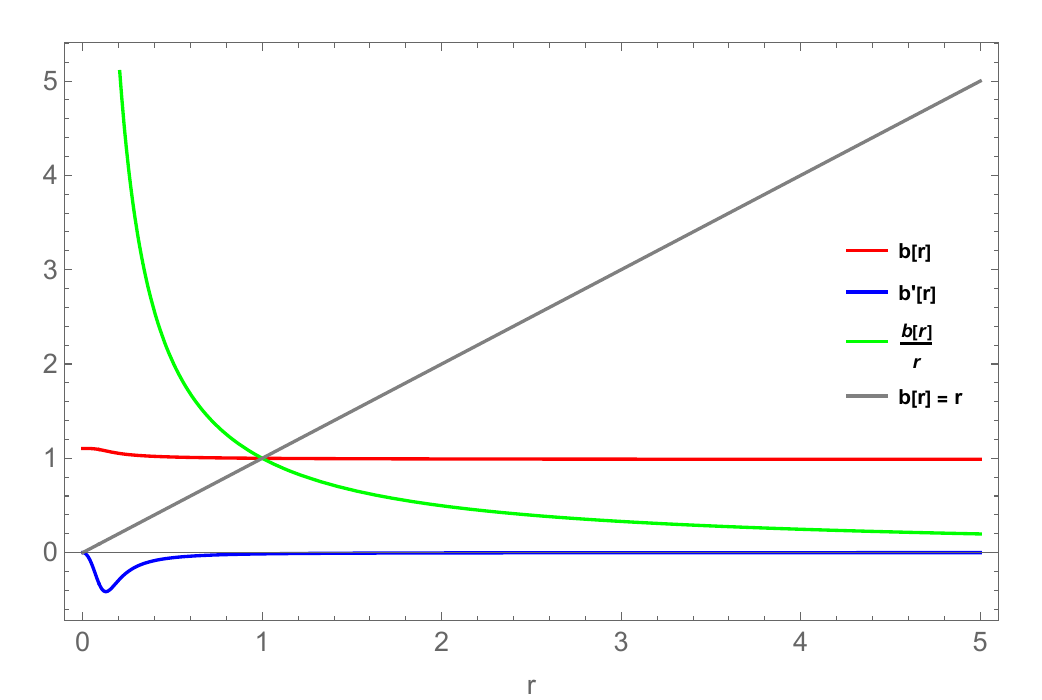}
    \caption{Plots for the $b(r)$, $b'(r)$, $b(r)/r$ and $b(r)=r$ for $\lambda=0.01$ and $r_{0}=1$.}
    \label{fig:shape_function}
\end{figure}
From Fig. \ref{fig:shape_function}, one can notice that the shape function \eqref{b(1)_2} satisfies the traversability conditions mentioned in the equations \eqref{SF_I} and \eqref{SF_III}.  
\begin{figure}[ht]
    \centering
    \includegraphics[width=0.4\textwidth]{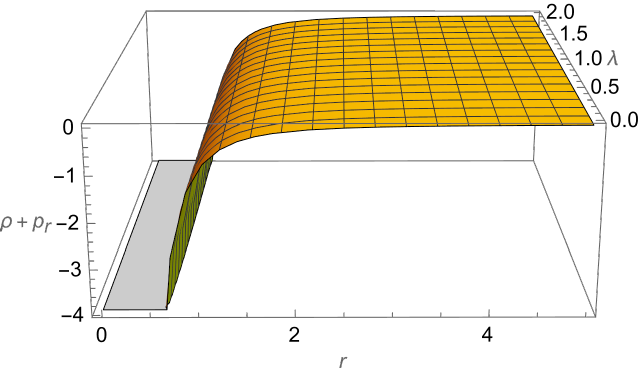}
    \caption{Plot for the $\rho+p_{r}$ for $r_{0}=1$ with $\lambda\in [0,2]$.}
    \label{fig:NEC1}
\end{figure}
\begin{figure}[ht]
    \centering
    \includegraphics[width=0.4\textwidth]{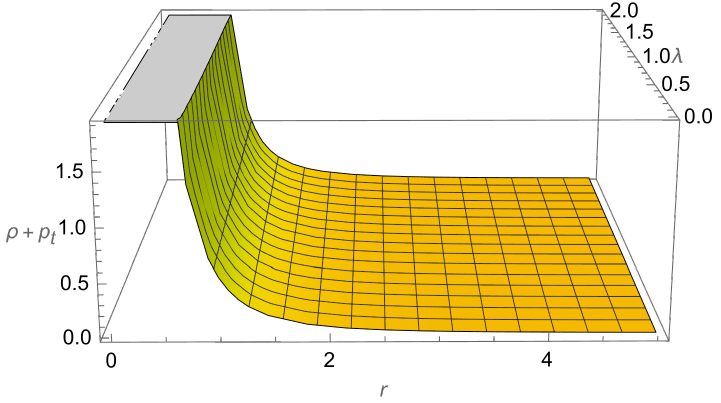}
    \caption{Plot for the $\rho+p_{t}$ for $r_{0}=1$ with $\lambda\in [0,2]$.}
    \label{fig:NEC2}
\end{figure}
\subsubsection{\texorpdfstring{$a$}{} as a constant}\label{caseII}

\begin{figure}
    \centering
    \includegraphics[width=0.4\textwidth]{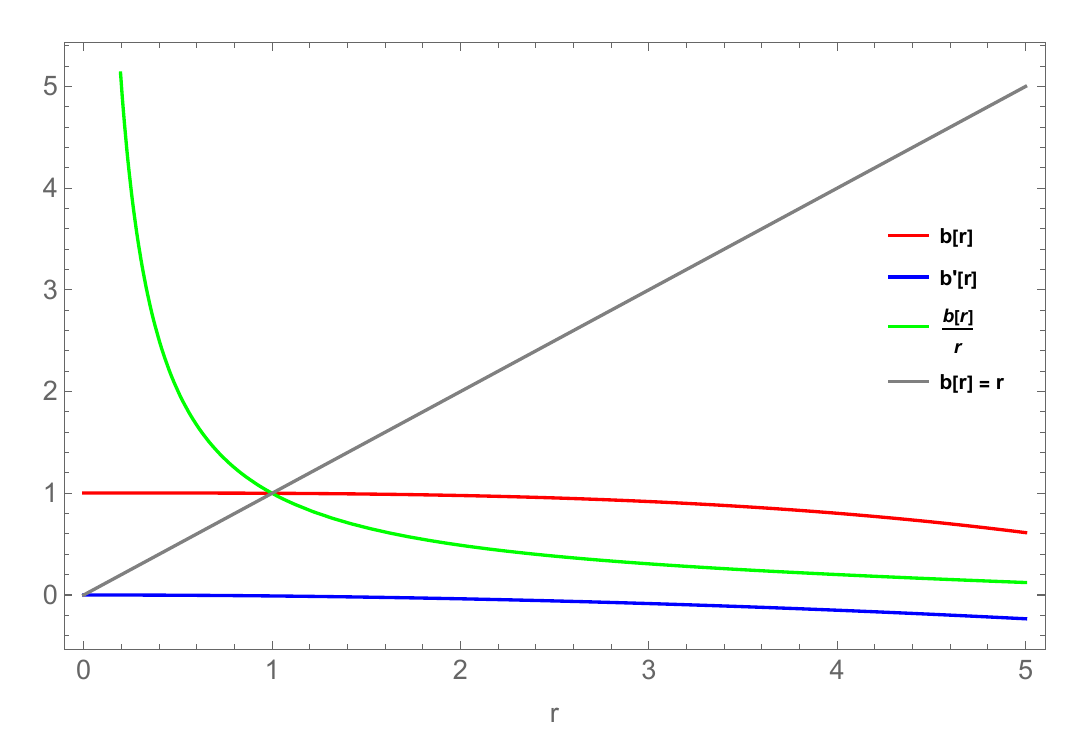}
    \caption{Plots for the $b(r)$, $b'(r)$, $b(r)/r$ and $b(r)=r$ for $\lambda=0.01$ and $r_{0}=1$.}
    \label{fig:shape_function_2}
\end{figure}

We can obtain the shape function of the Casimir wormhole from equations \eqref{rho1a} and \eqref{casimir_rho} for a constant value of plate distance $a$:

\begin{equation}
    b(r)=-\frac{\pi^{2} r^{3}}{6 \left(360 a^{4} +\pi^{2} \lambda \right)}+c_{2}\,.\label{SF2}
\end{equation}
$c_{2}$ is the integration constant, and it has a specific value
\begin{equation}
    c_{2}=\frac{\pi ^2 r_{0}^3}{6 \left(360 a^4+\pi ^2 \lambda \right)}+r_{0}\,.
\end{equation} 
The shape function \eqref{SF2} is derived using the energy density defined in Eq. \eqref{casimir_rho}, assuming that the distance between the plates remains constant. The primary objective is to investigate whether this constant distance impacts the wormhole scenario. Analyzing the behavior of the shape function and the energy conditions, it is observed that the derivative of the shape function, $b'(r)$, exhibits distinct behavior compared to other parameters, as illustrated in Figures \ref{fig:shape_function} and \ref{fig:shape_function_2}. Conversely, the energy conditions show minimal variation; however, the rate of energy flow decreases when the parameter $a$ is held constant, as opposed to when it is treated as a variable.
\begin{figure}
    \centering
    \includegraphics[width=0.4\textwidth]{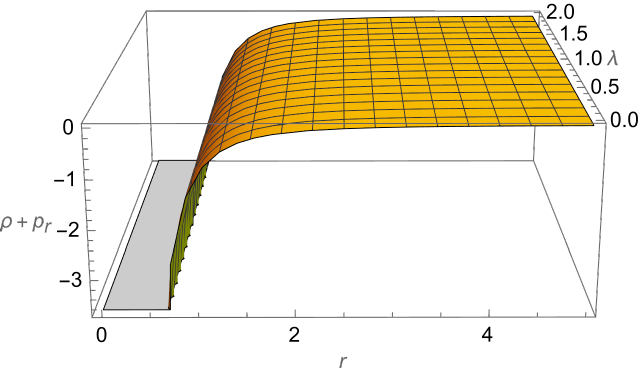}
    \caption{Plot for the $\rho+p_{r}$ for $r_{0}=1$ with $\lambda\in [0,2]$.}
    \label{fig:NEC_1a}
\end{figure}
\begin{figure}
    \centering
    \includegraphics[width=0.4\textwidth]{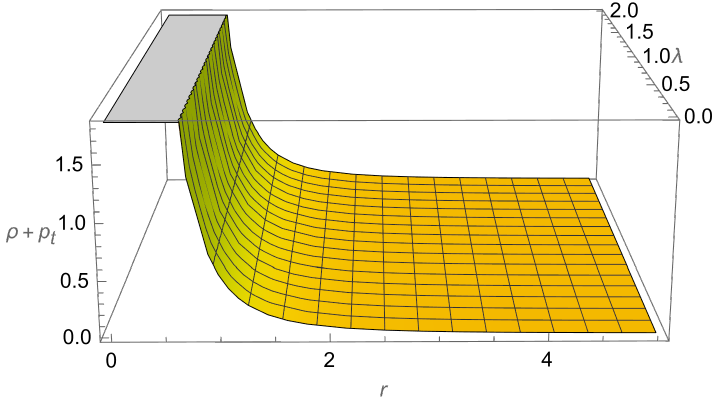}
    \caption{Plot for the $\rho+p_{t}$ for $r_{0}=1$ with $\lambda\in [0,2]$.}
    \label{fig:NEC_2a}
\end{figure}

 The NEC is a critical factor in analyzing the geometry and stability of a wormhole. In Einstein's GR, the NEC is typically violated in the context of wormholes, making it essential to examine this condition when studying traversable wormholes thoroughly. The NEC provides insight into the type of matter required to sustain a wormhole, often referred to as ``exotic matter,'' which possesses unusual properties. In the following discussion, we focus on the NEC to better understand the energy requirements and implications for the existence of traversable wormholes.

\paragraph*{Null Energy Conditions:} Wormhole spacetimes are supported by exotic matter, necessitating the specification of energy conditions for cases where the stress-energy tensor is diagonal (see equation \eqref{EMT_II}). When the radial pressure $p_{r}$ equals the tangential pressure $p_{t}$, the stress-energy tensor simplifies to that of a perfect fluid. While classical matter typically satisfies these energy conditions, they are known to be violated by certain quantum fields, such as those involved in the Casimir effect, highlighting the need for exotic matter in wormhole physics.

The NEC states that for any null vector $k_{\mu}$, $T_{\mu \nu}k^{\mu}k^{\nu}\geq 0$. In the case of a stress-energy tensor of the form \eqref{EMT_II}, we have
    \begin{equation}
        \forall ~~i\,, ~~~~\rho +p_{i}\geq 0\,.
    \end{equation}
    
In this study, we analyzed the NEC to assess the stability of the wormhole, focusing on both radial and tangential directions. Our findings indicate that the NEC is violated in the radial direction, regardless of whether the distance between the plates is constant or variable, as illustrated in Figures \ref{fig:NEC1} and \ref{fig:NEC_1a}. On the other hand, the NEC is satisfied in the tangential direction, with both cases displaying similar behavior, as shown in Figures \ref{fig:NEC2} and \ref{fig:NEC_2a}. These results provide crucial insights into the energy conditions governing the stability of the wormhole structure.
\subsection{The Casimir wormholes with GUP correction}\label{WithGUP}
Frassino and Panella applied the concepts of minimal length and the Generalized Uncertainty Principle (GUP) to calculate the finite energy between uncharged parallel plates. They derived the Hamiltonian and the corresponding corrections to the Casimir energy resulting from the introduction of minimal length. The Casimir energy was obtained for two distinct cases involving the construction of maximally localized states, with a first-order correction term in the minimal uncertainty parameter $\beta$ \cite{Frassino2012}.
\begin{equation}
    E_{i}=-\frac{\pi^{2}}{720}\frac{1}{a^{4}}\left(1+\zeta_{i}\frac{\beta}{a^{2}}\right)
\end{equation}
where 
\begin{subequations}
    \begin{eqnarray}
        \zeta_{KMM}&=&\pi^{2}\left(\frac{28+3\sqrt{10}}{14}\right)\,,\\
        \zeta_{DGS}&=&4\pi^{2}\left(\frac{3+\pi^{2}}{21}\right),
    \end{eqnarray}
    the suffix $i$ represents either the Kempf, Mangano, and Mann (KMM) \cite{Kempf1995} construction or the Detournay, Gabriel, and Spindel (DGS) \cite{Detournay2002} construction.
\end{subequations}
\begin{figure}[ht]
    \centering
    \includegraphics[width=1\linewidth]{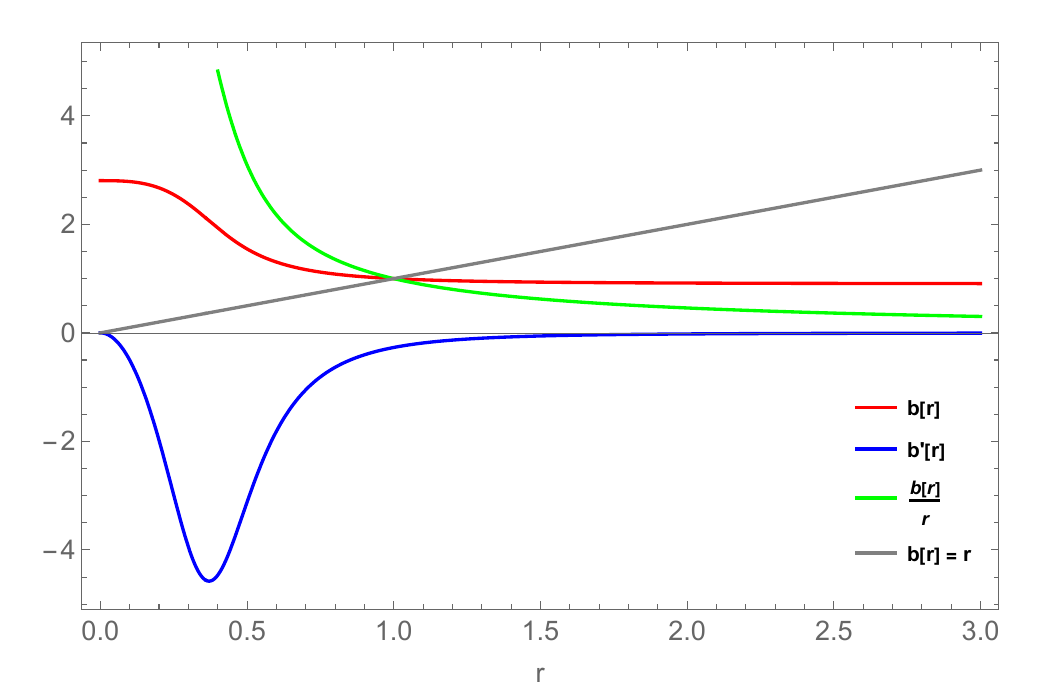}
    \includegraphics[width=1\linewidth]{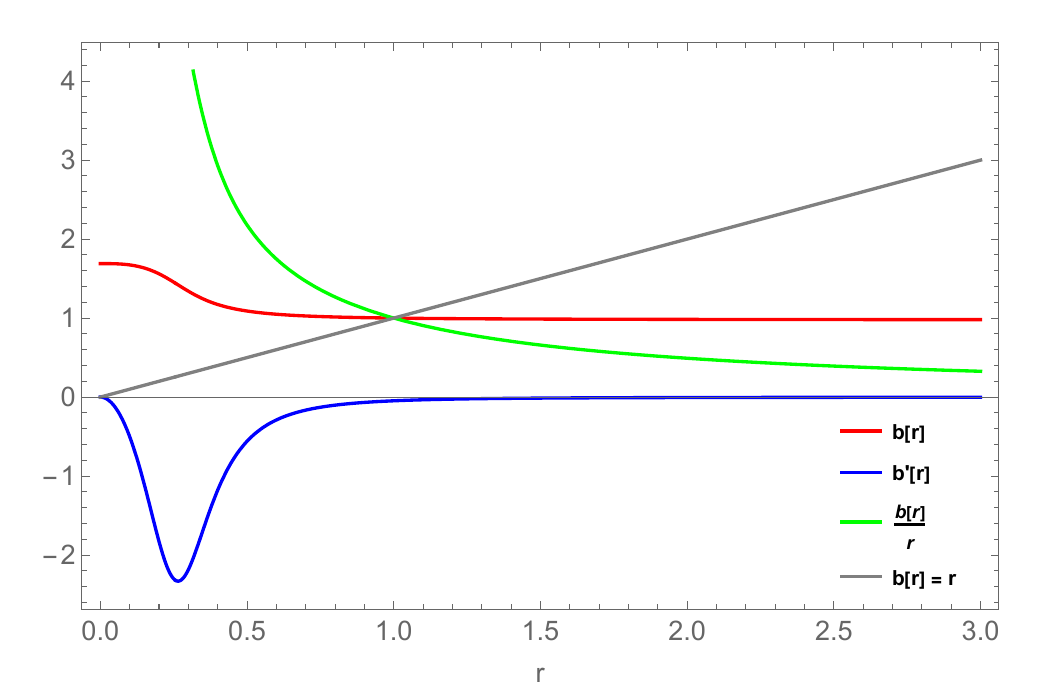}
    \caption{Plot for $b(r)$, $b'(r)$, $b(r)/r$ and $b(r)=r$ for $\lambda=0.01$, $\beta=0.1$ and $r_{0}=1$ for KKM (above) and DGS (below) construction.}
    \label{fig:enter-labela1}
\end{figure}
\begin{figure}[ht]
    \centering
    \includegraphics[width=1\linewidth]{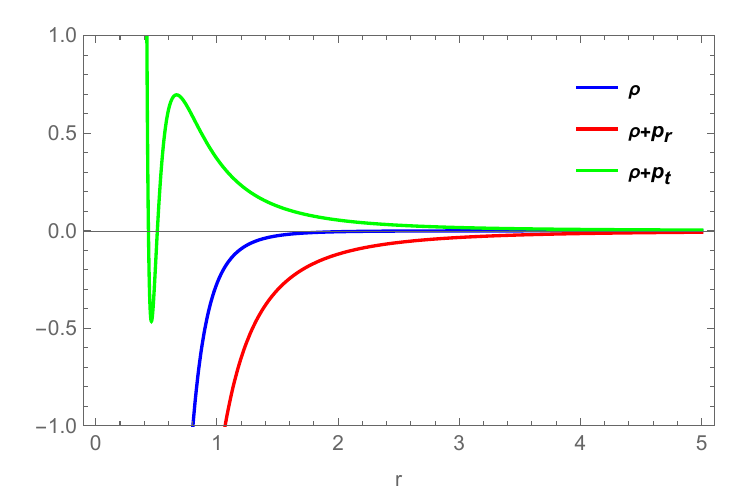}
    \includegraphics[width=1\linewidth]{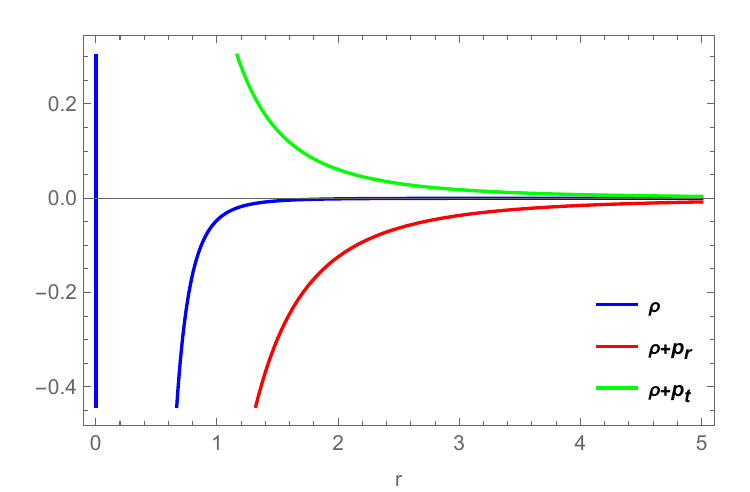}
    \caption{Plot for energy density and null energy conditions for KKM (above) and DGS (below) construction .}
    \label{fig:enter-labela2}
\end{figure}
Numerical solutions have been computed for both constructions, confirming the traversability of the model for specific parameter values. The GUP correction term is proportional to the minimal uncertainty parameter, $\beta$. As $\beta$ approaches zero, the shape function converges to that of the Casimir wormhole. Fig. \ref{fig:enter-labela1} illustrates that these shape functions satisfy the necessary conditions for traversability, including both the throat and flare-out conditions.

The behavior of the energy density and the NEC in both the radial and lateral directions for the KMM and DGS models is illustrated in Fig. \ref{fig:enter-labela2}. The graph shows that the NEC is violated in the radial direction due to the Casimir effect. However, for the lateral direction, the NEC violation occurs intermittently in the KMM model, whereas in the DGS model, the violation persists continuously throughout.

\section{Wormhole geometry}\label{WG}
Here, we will explain the $f(R,\mathcal{L}_{m})$ gravity wormhole geometry using an embedded diagram. We will also examine the shape function $b(r)$ from equation \eqref{b(r)_1}. We begin the analysis by considering the generic line element provided in Eq. \eqref{MTM}, which describes the spacetime geometry. Since we are working with spherically symmetric and static wormhole solutions, it is convenient to focus on the equatorial plane, where $\theta = \pi/2$. The symmetry of the problem justifies this simplification and allows us to reduce the complexity of the angular components. Specifically, by fixing $\theta$ to this value, the solid angle element, which is generally expressed as $d\Omega^{2} \equiv d\theta^{2} + (\sin\theta, d\phi)^{2}$, is reduced to $(d\phi)^{2}$. This approach streamlines the equations and facilitates the analysis of the wormhole geometry in a more tractable form.
\begin{equation}
    d\Omega^{2}=d\phi^{2}\,.
\end{equation}
Also, when we fix a constant time slice, i.e. $t = \text{constant}$, the line element \eqref{MTM} becomes
\begin{equation}
   ds^2=(1-b(r)/r)^{-1}dr^2+(rd\phi)^2  \,.
\end{equation}
It is practical to express the equatorial plane as a surface embedded in an Euclidean space using cylindrical coordinates
\begin{equation}
       ds^2=dz^{2}+dr^2+(rd\phi)^2  \,.
\end{equation}
which is straightforwardly rewritten as
\begin{equation}
    ds^{2}=\left[1+\left(\frac{dz}{dr}\right)^{2}\right]dr^{2}+(rd\phi)^2\,.
\end{equation}
\begin{equation}
    \frac{dz}{dr}=\pm \left[\frac{r}{b(r)}-1\right]^{-\frac{1}{2}}\,,\label{dz/dr}
\end{equation}
The embedded surface is defined by the equation $z=z(r)$. Notably, as $r$ approaches $+\infty$, the above formula yields
\begin{equation}
    \left.\frac{dz}{dr}\right|_{r \rightarrow +\infty}=0
\end{equation}
\begin{figure}[ht]
    \centering
    \includegraphics[width=0.4\textwidth]{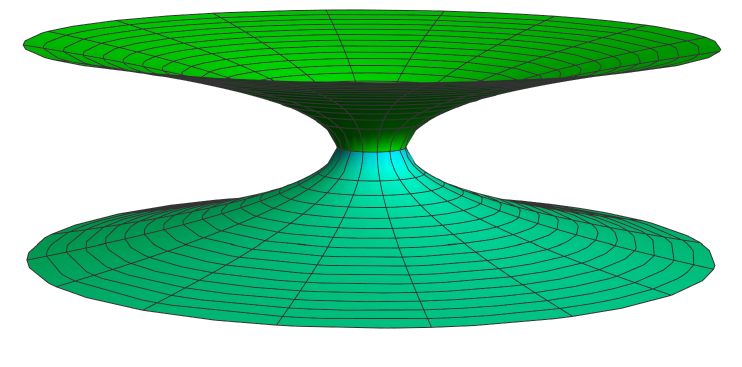}
    \caption{Wormhole geometry for the without GUP corrected wormhole.}
    \label{fig:wormhole_geometry}
\end{figure}
There are two asymptotically flat patches in the embedding diagram, highlighting distinct regions. It is important to note that an analytical integration of $b(r)$ using Eq. \eqref{dz/dr} is not feasible so, a numerical approach was employed to visualize the shape of the wormhole in Fig. \ref{fig:wormhole_geometry}. The definition of a wormhole implies that Eq. \eqref{dz/dr} diverges at $r = r_0$, indicating that the embedded surface becomes vertical at this point, corresponding to the throat of the wormhole. In this specific case, we assume $\lambda = 0.01$ for the embedded surface. The upper region of the universe, located above the throat, is depicted in green, while the lower region is represented in cyan. To construct the wormhole hypersurface, the curve $z(r)$ is rotated around a vertical axis. The radial distance is positive above the throat and negative below it. Additionally, it is noted that the embedded surface flattens as one moves far away from the throat, reflecting the asymptotic flatness of the geometry of the wormhole.

\begin{figure}
    \centering
    \includegraphics[width=0.8\linewidth]{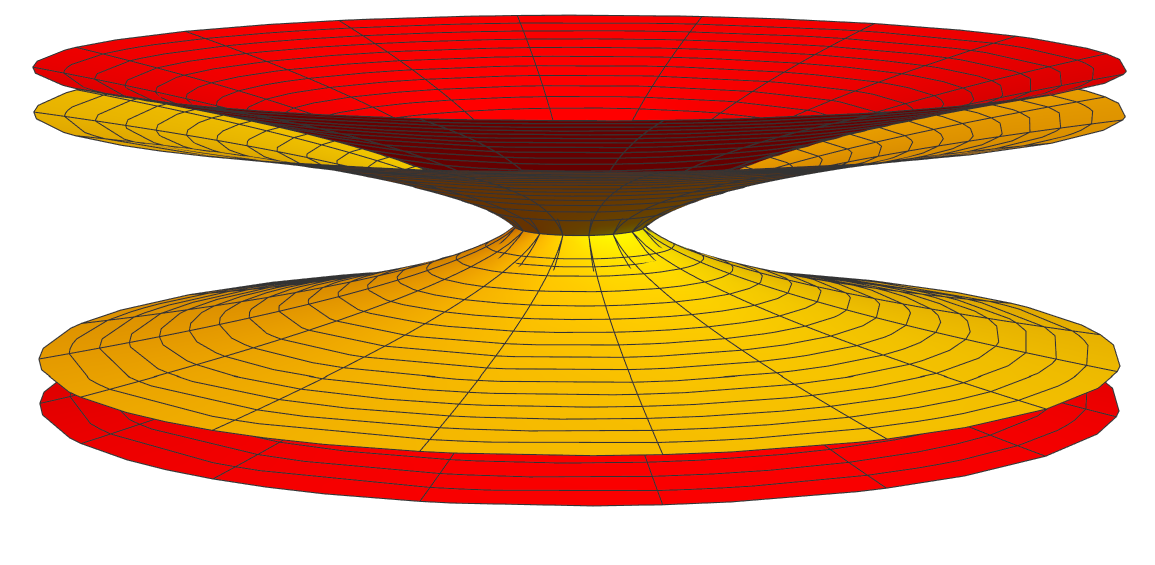}
    \includegraphics[width=0.8\linewidth]{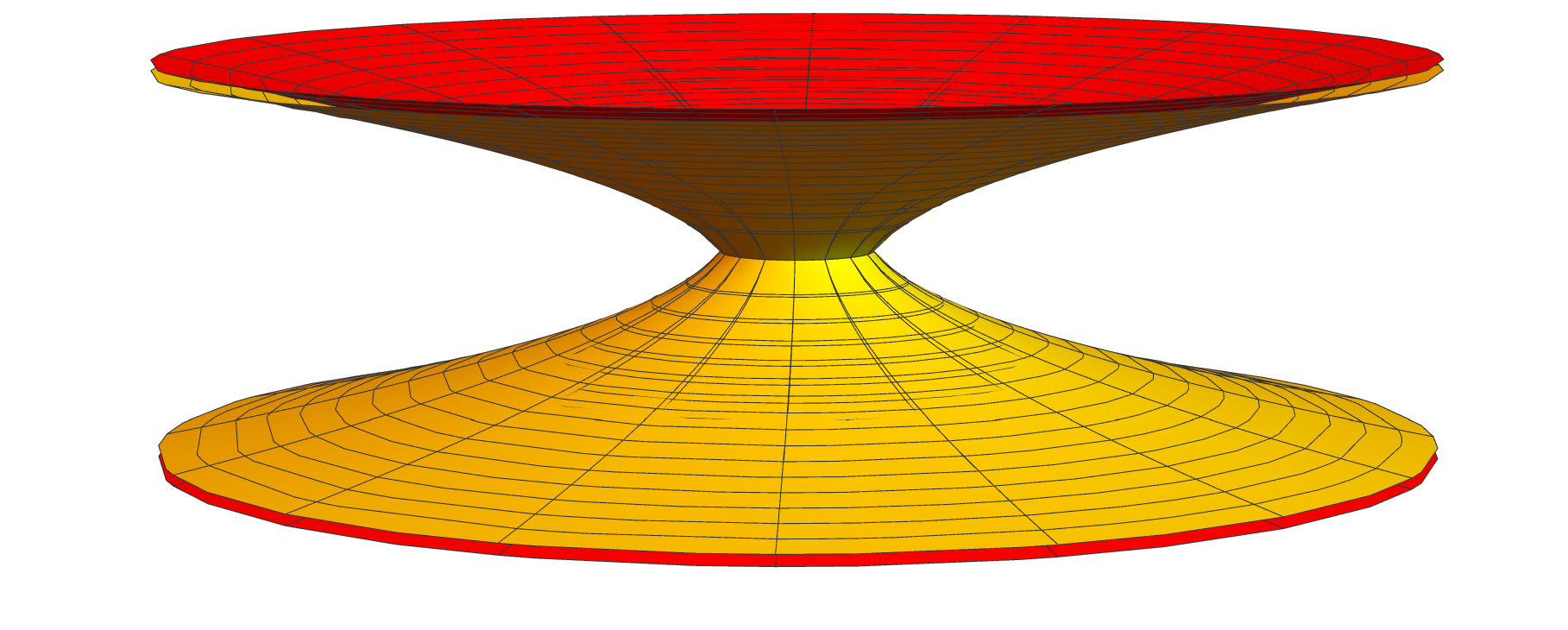}
    \caption{Comparison of wormhole geometries: The KMM construction (shaded in yellow) and the DGS construction (shaded in red). The upper plot corresponds to $\beta=0.5$, while the lower plot illustrates $\beta=0.1$.}
    \label{fig:enter-WHGUPGEOMETRY}
\end{figure}
In Fig. \ref{fig:enter-WHGUPGEOMETRY}, a noticeable difference between the geometries of the GUP-corrected Casimir wormholes with KMM and DGS  constructions can be seen. The wormhole with KMM  construction shows a flatter geometry compared to that with DGS construction. This is evident from the figures, where the surface of the wormhole with DGS construction is positioned above that of the KMM  construction in both positive and negative directions. This contrast underscores the differing effects of the GUP parameter on the curvature and overall shape of the wormhole geometries in these two models. Moreover, as the minimal uncertainty parameter approaches zero, the resulting geometry converges to that of a wormhole without GUP-corrected construction. This indicates that the influence of the GUP diminishes, and the geometry reverts to the classical wormhole structure, highlighting the role of the GUP in altering the spacetime configuration.

\section{Stability Condition}\label{SC}
In this section, we explore the stability of a zero-tidal-force wormhole by applying the equilibrium condition derived from the Tolman-Oppenheimer-Volkoff (TOV) equation \cite{Leon1993, Rahaman2014}. The TOV equation, typically employed to describe the balance of forces in stellar structures, provides a framework for analyzing the equilibrium between gravitational forces and the internal pressure of the matter composing the wormhole. By enforcing the zero-tidal-force condition, we restrict our investigation to scenarios where tidal effects are absent, allowing for a more focused examination of the structural stability of the wormhole. This analysis is essential for assessing whether the wormhole can remain stable under small perturbations, ensuring its traversability without requiring exotic matter that violates physical energy conditions.
\begin{equation}
    \frac{d p_{r}}{d\rho}+\Phi '(\rho +p_{r})+\frac{2}{r}(p_{r}-p_{t})=0\,, \label{cond_eq} 
\end{equation}
The equilibrium state of a structure is governed by three primary forces: gravitational force ($F_{g}$), hydro-static force ($F_{h}$), and the anisotropic forces ($F_{a}$) acting on the system. These forces can written as follows
\begin{subequations}
    \begin{eqnarray}
            F_{g}&=&-\Phi '(\rho +p_{r})\,,\\
            F_{h}&=&-\frac{d p_{r}}{dr}\,,\\
            F_{a}&=&\frac{2(p_{t}-p_{r})}{r}\,.
    \end{eqnarray}
\end{subequations}
In a Morris-Thorne wormhole, the anisotropic pressure plays a role in shaping the equilibrium condition, as derived from Eq. \eqref{cond_eq}. This result leads the fundamental relationship $F_{g} + F_{h} + F_{a} = 0$, where the balance is crucial for ensuring the stability of the wormhole.

Since $\Phi' \equiv 0$, the equilibrium condition simplifies to
\begin{equation}
   F_{h}+F_{a}=0.
\end{equation}
Based on the preceding analysis, the hydro-static force and anisotropic forces are determined as follows,
\begin{subequations}
    \begin{eqnarray}
        F_{h}&=&\frac{r \left(3 b'(r)-r b''(r)\right)-3 b(r)}{r^4 (\lambda  R+1)^2}\,,\\
        F_{a}&=&\frac{3 b(r)-r b'(r)}{r^4 (\lambda  R+1)^2}\,.
    \end{eqnarray}
\end{subequations}
From the above conditions, one can write
\begin{equation}
    F_{h}+F_{a}=\frac{2 b'(r)-r b''(r)}{r^3 (\lambda  R+1)^2}
\end{equation}

\begin{figure}[ht]
    \centering
    \includegraphics[width=1\linewidth]{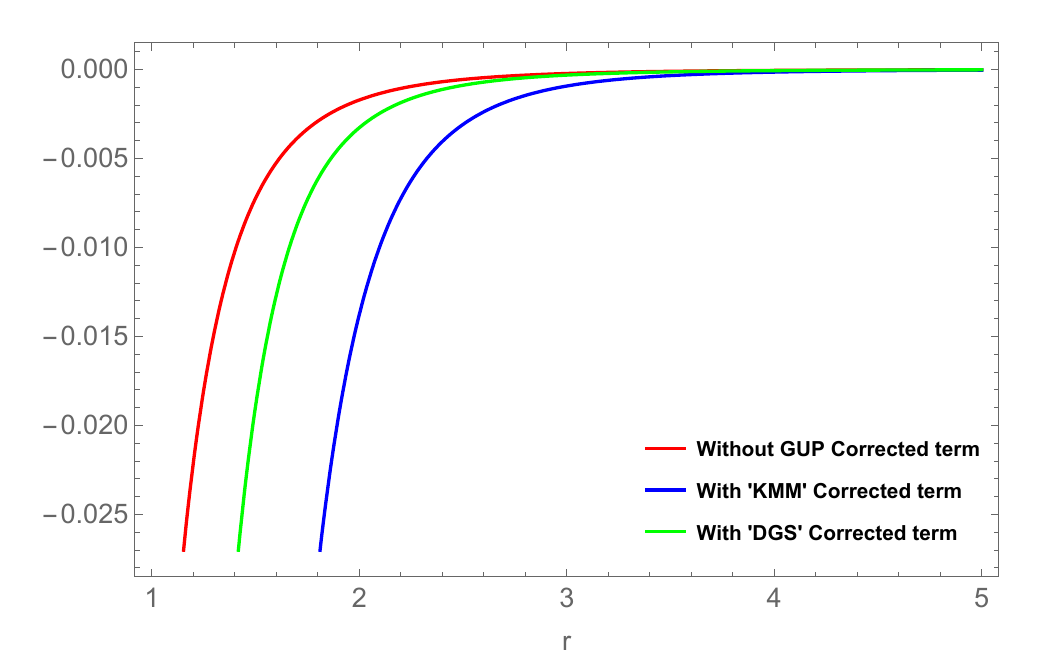}
    \caption{Comparison of different equilibrium conditions in the absence and presence of GUP corrections. The red curve represents the equilibrium condition without GUP correction. The blue and green curves correspond to the GUP-corrected terms based on the KMM and DGS constructions, respectively.}
    \label{fig:stability}
\end{figure}

A comparison of equilibrium conditions with and without GUP corrections is shown in FIG. \ref{fig:stability}. The red curve represents the equilibrium condition without GUP correction. In contrast, the blue and green curves represent the GUP-corrected equilibrium conditions based on the KMM model and the DGS model, respectively. The plot reveals that the equilibrium condition without GUP correction converges to zero more quickly than the GUP-corrected cases. Furthermore, the DGS-corrected curve converges to zero faster than the KMM-corrected curve, likely due to variations in the scalar values of the two models.

\section{Conclusion}\label{Conclusion}
In this paper, we have investigated the traversable wormholes within the framework of the modified gravity model $f(R,\mathcal{L}_{m})=R/2+(1+\lambda R)\mathcal{L}_{m}$, incorporating the effect of the Casimir force. By analyzing cases where the distance between the plates, $a$, is treated as both a variable and a constant, we have derived the necessary conditions for wormhole traversability. Our findings show that while the NEC is satisfied laterally in both cases, it is violated in the radial direction. The violation is attributed to the influence of the Casimir effect, which plays a key role in shaping our analysis. The Casimir effect, a quantum phenomenon arising from the vacuum fluctuations between closely spaced conductive plates, significantly affects the energy conditions within the wormhole. By incorporating the Casimir effect into our framework, we gain a deeper understanding of how quantum forces can alter the behavior and stability of traversable wormholes, particularly in the context of modified gravity theories. This underscores the importance of considering quantum effects in analyzing exotic geometries and their traversability. These results highlight the intriguing impact of quantum effects, such as the Casimir effect, on the behavior of traversable wormholes, particularly within the context of modified gravity theories. 

The investigation of wormhole traversability and stability, both with and without GUP corrections, has provided valuable insights into the behavior of these exotic structures. The analysis revealed key differences in the energy conditions and equilibrium states between the corrected and uncorrected models. In particular, the incorporation of GUP corrections, as seen in the KMM and DGS frameworks, led to distinct variations in the stability and traversability of the wormholes. These corrections influenced the conditions $\rho + p_{r} < 0$ and $\rho < 0$ for $r > 0$, highlighting the role of quantum effects, such as the Casimir effect, in shaping the properties of wormholes. Furthermore, the stability analysis showed that wormhole geometries tend to flatten at larger radial distances, especially in models without GUP corrections. The convergence patterns in the equilibrium plots further underscore the importance of quantum gravity modifications in understanding the stability and dynamics of traversable wormholes.  These findings emphasize the intricate relation between quantum corrections and classical wormhole geometries, offering a deeper understanding of their stability and traversability.

\section*{Acknowledgement} BM acknowledges the support of Council of Scientific and Industrial Research (CSIR) for the project grant (No. 03/1493/23/EMR II).

\section*{Reference}
\bibliographystyle{utphys}

\bibliography{main}
\end{document}